\RequirePackage{lineno}
\documentclass[aps,prd,twocolumn,showpacs,superscriptaddress,groupedaddress]{revtex4}

\usepackage{graphicx}

\usepackage{dcolumn}
\usepackage{bm}

\usepackage{textcomp} 
\usepackage{amstext} 

\usepackage[utf8]{inputenc}
\usepackage{graphicx}
\usepackage{epsfig}

\begin{document}


\mbox{{\em Published in Phys. Rev. D 102, 072005} \hspace{6.8cm} FERMILAB-PUB-20-363-E}




%
\vspace*{-0.1cm}

\title{Studies of {\boldmath  $X(3872)$} and {\boldmath  $\psi(2S)$} production in {\boldmath  $p\bar{p}$} collisions at 1.96 TeV }

%
\affiliation{LAFEX, Centro Brasileiro de Pesquisas F\'{i}sicas, Rio de Janeiro, RJ 22290, Brazil}
\affiliation{Universidade do Estado do Rio de Janeiro, Rio de Janeiro, RJ 20550, Brazil}
\affiliation{Universidade Federal do ABC, Santo Andr\'e, SP 09210, Brazil}
\affiliation{University of Science and Technology of China, Hefei 230026, People's Republic of China}
\affiliation{Universidad de los Andes, Bogot\'a, 111711, Colombia}
\affiliation{Charles University, Faculty of Mathematics and Physics, Center for Particle Physics, 116 36 Prague 1, Czech Republic}
\affiliation{Czech Technical University in Prague, 116 36 Prague 6, Czech Republic}
\affiliation{Institute of Physics, Academy of Sciences of the Czech Republic, 182 21 Prague, Czech Republic}
\affiliation{Universidad San Francisco de Quito, Quito 170157, Ecuador}
\affiliation{LPC, Universit\'e Blaise Pascal, CNRS/IN2P3, Clermont, F-63178 Aubi\`ere Cedex, France}
\affiliation{LPSC, Universit\'e Joseph Fourier Grenoble 1, CNRS/IN2P3, Institut National Polytechnique de Grenoble, F-38026 Grenoble Cedex, France}
\affiliation{CPPM, Aix-Marseille Universit\'e, CNRS/IN2P3, F-13288 Marseille Cedex 09, France}
\affiliation{LAL, Univ. Paris-Sud, CNRS/IN2P3, Universit\'e Paris-Saclay, F-91898 Orsay Cedex, France}
\affiliation{LPNHE, Universit\'es Paris VI and VII, CNRS/IN2P3, F-75005 Paris, France}
\affiliation{IRFU, CEA, Universit\'e Paris-Saclay, F-91191 Gif-Sur-Yvette, France}
\affiliation{IPHC, Universit\'e de Strasbourg, CNRS/IN2P3, F-67037 Strasbourg, France}
\affiliation{IPNL, Universit\'e Lyon 1, CNRS/IN2P3, F-69622 Villeurbanne Cedex, France and Universit\'e de Lyon, F-69361 Lyon CEDEX 07, France}
\affiliation{III. Physikalisches Institut A, RWTH Aachen University, 52056 Aachen, Germany}
\affiliation{Physikalisches Institut, Universit\"at Freiburg, 79085 Freiburg, Germany}
\affiliation{II. Physikalisches Institut, Georg-August-Universit\"at G\"ottingen, 37073 G\"ottingen, Germany}
\affiliation{Institut f\"ur Physik, Universit\"at Mainz, 55099 Mainz, Germany}
\affiliation{Ludwig-Maximilians-Universit\"at M\"unchen, 80539 M\"unchen, Germany}
\affiliation{Panjab University, Chandigarh 160014, India}
\affiliation{Delhi University, Delhi-110 007, India}
\affiliation{Tata Institute of Fundamental Research, Mumbai-400 005, India}
\affiliation{University College Dublin, Dublin 4, Ireland}
\affiliation{Korea Detector Laboratory, Korea University, Seoul, 02841, Korea}
\affiliation{CINVESTAV, Mexico City 07360, Mexico}
\affiliation{Nikhef, Science Park, 1098 XG Amsterdam, the Netherlands}
\affiliation{Radboud University Nijmegen, 6525 AJ Nijmegen, the Netherlands}
\affiliation{Joint Institute for Nuclear Research, Dubna 141980, Russia}
\affiliation{Institute for Theoretical and Experimental Physics, Moscow 117259, Russia}
\affiliation{Moscow State University, Moscow 119991, Russia}
\affiliation{Institute for High Energy Physics, Protvino, Moscow region 142281, Russia}
\affiliation{Petersburg Nuclear Physics Institute, St. Petersburg 188300, Russia}
\affiliation{Instituci\'{o} Catalana de Recerca i Estudis Avan\c{c}ats (ICREA) and Institut de F\'{i}sica d'Altes Energies (IFAE), 08193 Bellaterra (Barcelona), Spain}
\affiliation{Uppsala University, 751 05 Uppsala, Sweden}
\affiliation{Taras Shevchenko National University of Kyiv, Kiev, 01601, Ukraine}
\affiliation{Lancaster University, Lancaster LA1 4YB, United Kingdom}
\affiliation{Imperial College London, London SW7 2AZ, United Kingdom}
\affiliation{The University of Manchester, Manchester M13 9PL, United Kingdom}
\affiliation{University of Arizona, Tucson, Arizona 85721, USA}
\affiliation{University of California Riverside, Riverside, California 92521, USA}
\affiliation{Florida State University, Tallahassee, Florida 32306, USA}
\affiliation{Fermi National Accelerator Laboratory, Batavia, Illinois 60510, USA}
\affiliation{University of Illinois at Chicago, Chicago, Illinois 60607, USA}
\affiliation{Northern Illinois University, DeKalb, Illinois 60115, USA}
\affiliation{Northwestern University, Evanston, Illinois 60208, USA}
\affiliation{Indiana University, Bloomington, Indiana 47405, USA}
\affiliation{Purdue University Calumet, Hammond, Indiana 46323, USA}
\affiliation{University of Notre Dame, Notre Dame, Indiana 46556, USA}
\affiliation{Iowa State University, Ames, Iowa 50011, USA}
\affiliation{University of Kansas, Lawrence, Kansas 66045, USA}
\affiliation{Louisiana Tech University, Ruston, Louisiana 71272, USA}
\affiliation{Northeastern University, Boston, Massachusetts 02115, USA}
\affiliation{University of Michigan, Ann Arbor, Michigan 48109, USA}
\affiliation{Michigan State University, East Lansing, Michigan 48824, USA}
\affiliation{University of Mississippi, University, Mississippi 38677, USA}
\affiliation{University of Nebraska, Lincoln, Nebraska 68588, USA}
\affiliation{Rutgers University, Piscataway, New Jersey 08855, USA}
\affiliation{Princeton University, Princeton, New Jersey 08544, USA}
\affiliation{State University of New York, Buffalo, New York 14260, USA}
\affiliation{University of Rochester, Rochester, New York 14627, USA}
\affiliation{State University of New York, Stony Brook, New York 11794, USA}
\affiliation{Brookhaven National Laboratory, Upton, New York 11973, USA}
\affiliation{Langston University, Langston, Oklahoma 73050, USA}
\affiliation{University of Oklahoma, Norman, Oklahoma 73019, USA}
\affiliation{Oklahoma State University, Stillwater, Oklahoma 74078, USA}
\affiliation{Oregon State University, Corvallis, Oregon 97331, USA}
\affiliation{Brown University, Providence, Rhode Island 02912, USA}
\affiliation{University of Texas, Arlington, Texas 76019, USA}
\affiliation{Southern Methodist University, Dallas, Texas 75275, USA}
\affiliation{Rice University, Houston, Texas 77005, USA}
\affiliation{University of Virginia, Charlottesville, Virginia 22904, USA}
\affiliation{University of Washington, Seattle, Washington 98195, USA}
\author{V.M.~Abazov} \affiliation{Joint Institute for Nuclear Research, Dubna 141980, Russia}
\author{B.~Abbott} \affiliation{University of Oklahoma, Norman, Oklahoma 73019, USA}
\author{B.S.~Acharya} \affiliation{Tata Institute of Fundamental Research, Mumbai-400 005, India}
\author{M.~Adams} \affiliation{University of Illinois at Chicago, Chicago, Illinois 60607, USA}
\author{T.~Adams} \affiliation{Florida State University, Tallahassee, Florida 32306, USA}
\author{J.P.~Agnew} \affiliation{The University of Manchester, Manchester M13 9PL, United Kingdom}
\author{G.D.~Alexeev} \affiliation{Joint Institute for Nuclear Research, Dubna 141980, Russia}
\author{G.~Alkhazov} \affiliation{Petersburg Nuclear Physics Institute, St. Petersburg 188300, Russia}
\author{A.~Alton$^{a}$} \affiliation{University of Michigan, Ann Arbor, Michigan 48109, USA}
\author{A.~Askew} \affiliation{Florida State University, Tallahassee, Florida 32306, USA}
% ...
%
%
\collaboration{The D0 Collaboration\footnote{with visitors from
$^{a}$Augustana College, Sioux Falls, SD 57197, USA,
$^{b}$The University of Liverpool, Liverpool L69 3BX, UK,
$^{c}$Deutshes Elektronen-Synchrotron (DESY), Notkestrasse 85, Germany,
$^{d}$CONACyT, M-03940 Mexico City, Mexico,
$^{e}$SLAC, Menlo Park, CA 94025, USA,
$^{f}$University College London, London WC1E 6BT, UK,
$^{g}$Centro de Investigacion en Computacion - IPN, CP 07738 Mexico City, Mexico,
$^{h}$Universidade Estadual Paulista, S\~ao Paulo, SP 01140, Brazil,
$^{i}$Karlsruher Institut f\"ur Technologie (KIT) - Steinbuch Centre for Computing (SCC),
D-76128 Karlsruhe, Germany,
$^{j}$Office of Science, U.S. Department of Energy, Washington, D.C. 20585, USA,
$^{l}$Kiev Institute for Nuclear Research (KINR), Kyiv 03680, Ukraine,
$^{m}$University of Maryland, College Park, MD 20742, USA,
$^{n}$European Orgnaization for Nuclear Research (CERN), CH-1211 Geneva, Switzerland,
$^{o}$Purdue University, West Lafayette, IN 47907, USA,
$^{p}$Institute of Physics, Belgrade, Belgrade, Serbia,
and
$^{q}$P.N. Lebedev Physical Institute of the Russian Academy of Sciences, 119991, Moscow, Russia.
$^{\ddag}$Deceased.
}} \noaffiliation
\vskip 0.25cm





\date{\today}

\begin{abstract}
\vspace*{0.1cm}
We  present various properties of the production of the $X(3872)$
and $\psi(2S)$ states based on \mbox{10.4 fb$^{-1}$} collected by the D0 experiment in Tevatron $p \bar p$ collisions at $\sqrt{s}$ = 1.96~TeV.
For both states, we measure the nonprompt fraction $f_{NP}$  of the inclusive production rate
due to decays of  $b$-flavored hadrons.
 We find the  $f_{NP}$  values systematically below those
 obtained at the LHC. The  $f_{NP}$ fraction 
for $\psi(2S)$ increases
with transverse momentum, whereas for the $X(3872)$ it is constant  within large uncertainties, in agreement with the LHC results.
The ratio of prompt to nonprompt $\psi(2S)$ production, $(1 - f_{NP}) / f_{NP}$, decreases only slightly
going from the Tevatron to the LHC, but for the $X(3872)$, this ratio decreases by a factor of about 3.
We test the soft-pion signature
 of the $X(3872)$ modeled
as a weakly bound charm-meson pair by studying the production of the $X(3872)$
as a function of the kinetic energy of the $X(3872)$ and the pion
in the  $X(3872)\pi$ center-of-mass frame.
For a subsample consistent with prompt production, the results
are incompatible with a strong enhancement in the  production of the $X(3872)$ at the small
kinetic  energy of the $X(3872)$ and the $\pi$ in the $X(3872)\pi$ center-of-mass frame
expected for the $X$+soft-pion production mechanism.
For events consistent with being due to decays of $b$ hadrons,
there is no significant evidence for the soft-pion effect, but its presence at the level
expected for the binding energy of 0.17~MeV and the momentum scale $\Lambda=M(\pi)$ is not ruled out.
\end{abstract}



\maketitle




\section{\label{sec:intro}Introduction}

Fifteen years after the discovery of the state $X(3872)$~\cite{belle}
(also named $\chi_{c1}(3872)$  \cite{pdg})
its nature is still debated. Its proximity to the $D^0 \bar D^{*0}$
threshold suggests a  charm-meson molecule loosely
bound by the pion exchange potential, first suggested by Tornqvist~\cite{tornqvist}.
The molecular model also explains the isospin breaking decay to $J/\psi \rho$
that is not allowed for a pure charmonium state.
However, the copious prompt production of the  $X(3872)$ at hadron colliders
has been  used as an argument against a pure molecule interpretation~\cite{amol}.
 With the binding energy  less than  1~MeV, the average  distance between
the two components is a few femtometers.
It has been argued that the production of such an extended object in the hadron collision
environment is strongly disfavored and  is better described by a compact
charm-anticharm or diquark-antidiquark structure.
Meng, Gao and Chao~\cite{meng2005} proposed that the $X(3872)$ is a mixture of
the conventional charmonium state $\chi_{c1}(2P)$ and a $D^0 \bar D^{*0}$ molecule.
In this picture, the short-distance production proceeds through the
 $\chi_{c1}(2P)$ component, while the  $D^0 \bar D^{*0}$
component is responsible for hadronic decays.
An evaluation of the production cross section of the $X(3872)$~\cite{meng2017} through its  $\chi_{c1}(2P)$ component 
gives a good description of
the differential cross section for the prompt production of $X(3872)$ measured
by CMS~\cite{cms} and ATLAS~\cite{8tev3872}.

Recently, Braaten $et al.$~\cite{bra3872,bra3872b} have revised the calculation of
the production of the $X(3872$) under the purely molecular hypothesis by
taking into account the
formation of $D^* \bar D^*$ at short distances followed by the 
rescattering of the charm mesons onto $X\pi$. According to the authors,
such a process should
 be easily observable at hadron colliders as an increased
event rate at small values of the kinetic energy $T(X\pi)$ of the $X(3872)$ and
the ``soft'' pion  in the $X(3872)\pi$ center-of-mass frame  and should provide a clean test
of the molecular structure of the $X(3872)$.

In this article, we present production properties of  the $X(3872)$
in Tevatron $p \bar p$ collisions  at the energy $\sqrt{s}$ = 1.96~TeV  and compare them with those of the conventional  charmonium
state $\psi(2S)$. Section \ref{sec:event} describes relevant experimental details and the event selections. 
In Sec.  \ref{sec:4tracks}, we present the transverse momentum $p_T$ and pseudorapidity $\eta$ dependence
of the fraction $f_{NP}$ of the inclusive production rate due to nonprompt decays of
$b$-flavored hadrons.
In Sec. \ref{sec:5tracks}, we study the hadronic activity around the  $X(3872)$ and  $\psi(2S)$.
We also test the soft-pion signature of the $X(3872)$ as a weakly bound
charm-meson pair by studying the production of $X(3872)$ plus a comoving pion at small  $T(X\pi)$.
 As a control process, we use the production
of the charmonium state $\psi(2S)$, for which this 
production mechanism does not apply. We summarize the findings in Sec. \ref{sec:sum}.

\section{\label{sec:event}The D0 detector, event reconstruction  and selection}

The D0 detector has a central tracking system consisting of a silicon
microstrip tracker  and the central fiber tracker, both located within a
1.9~T superconducting solenoidal magnet~\cite{d0det, layer0}. A muon
system, covering the pseudorapidity interval $|\eta|<2$~\cite{eta}, consists of a layer of tracking
detectors and scintillation trigger counters in front of 1.8~T iron toroidal
magnets, followed by two similar layers after the toroids~\cite{run2muon}. 
Events used in this analysis are collected with both single-muon and dimuon triggers.
Single-muon triggers require a coincidence of signals in trigger elements inside and outside
the toroidal magnets.
 All dimuon triggers  require at least one muon to have track segments  after the toroid;
muons in the forward region are always required to penetrate the   toroid.
The minimum muon transverse momentum is 1.5 GeV.
No minimum $p_T$ requirement is applied to the muon pair,
but the effective threshold is approximately 4 GeV due to the requirement for muons
to penetrate the toroids,  and the average  value for
accepted events is 10 GeV.

We select two samples, referred to as  4-track  and  5-track selections.
To select   4-track candidates, we reconstruct $J/\psi \rightarrow \mu^+ \mu^-$
 decay candidates accompanied by two particles of opposite charge assumed to be pions, 
 with transverse momentum $p_T$ with respect to the beam axis 
greater than 0.5~GeV.
We perform a kinematic fit  under the hypothesis that the muons come from
the $J/\psi$, and that the $J/\psi$ and the two particles  originate from
the same space point.
In the fit,  the dimuon invariant mass is constrained to the world average value
of the  $J/\psi$ meson  mass~\cite{pdg}.
The  track parameters ($p_T$ and position and direction in three dimensions) readjusted
according to the  fit
are used in the calculation of the invariant mass $M(J/\psi \pi^+\pi^-)$ and the decay length vector ${\vec L_{\rm xy}}$,
which is the transverse projection of the vector directed from the primary vertex to the $J/\psi \pi^+ \pi^-$ production vertex.
The two-pion mass for each accepted $J/\psi \pi^+ \pi^-$ candidate is required
to be greater than 0.35~GeV (0.5~GeV) for $\psi(2S)$ ($X(3872)$) candidates.
These conditions have a signal acceptance of more than 99\% 
while reducing the combinatorial background.
The transverse momentum of the  $J/\psi \pi^+ \pi^-$  system is required to be greater than 7~GeV.
All tracks in a given event are considered, and all combinations of tracks satisfying the conditions
stated are kept.
The mass windows $3.62 < M(J/\psi \pi^+ \pi^-) < 3.78$~GeV and $3.75 < M(J/\psi \pi^+ \pi^-) < 4.0$~GeV
are used for $\psi(2S)$ and $X(3872)$ selections, respectively. The rates of multiple entries within these ranges are less than 10\%.

\begin{figure}[htb]
\includegraphics[scale=0.4]{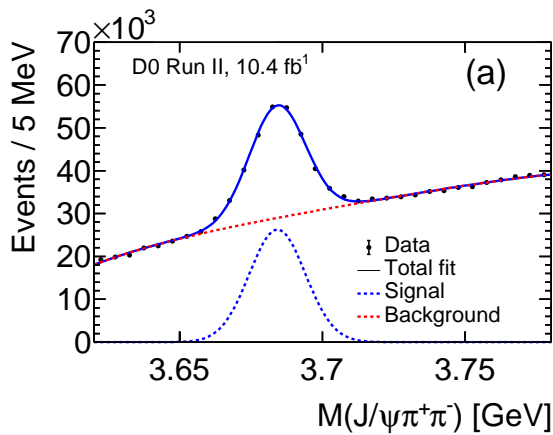}
\includegraphics[scale=0.4]{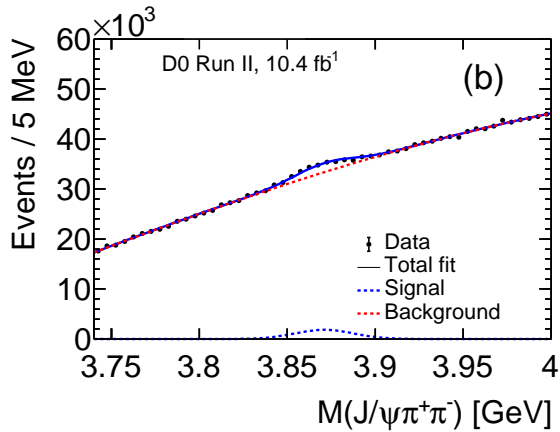}
\caption{\label{fig:mxfit} The invariant mass  $M(J/\psi \pi^+  \pi^-)$
for  (a)  the $\psi(2S)$  and  (b)
the $X(3872)$  selection criteria for the 4-track selection.
}
\end{figure}

Fits to the $M(J/\psi \pi^+  \pi^-)$ distribution for the 4-track selection
are shown in Fig.~\ref{fig:mxfit}.
In the  fits, the signal is modeled by a Gaussian function with a free mass and width.
Background is described by a fourth-order Chebyshev  polynomial.
The fits yield  $126\,891\,\pm\,770$ and  $16\,423\,\pm\,1031$  events of  $\psi(2S)$ and $X(3872)$, with
mass parameters of 3684.88$\,\pm\,$0.07~MeV and 3871.0$\,\pm\,$0.2~MeV, and 
mass resolutions of  9.7$\,\pm\,$0.1~MeV  and 16.7$\,\pm\,$0.9~MeV, respectively.
These mass resolutions are used in all subsequent fits.

\begin{figure}[htb]
\includegraphics[scale=0.4]{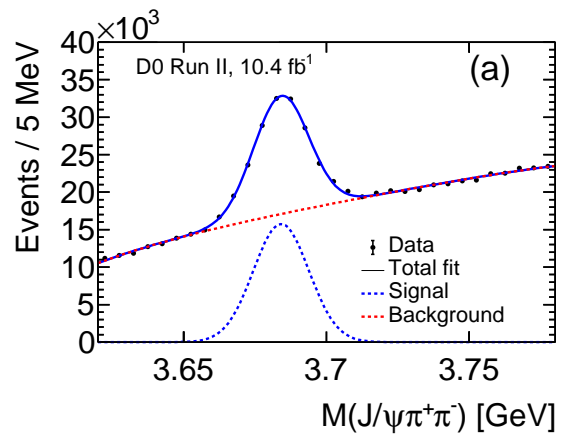}
\includegraphics[scale=0.4]{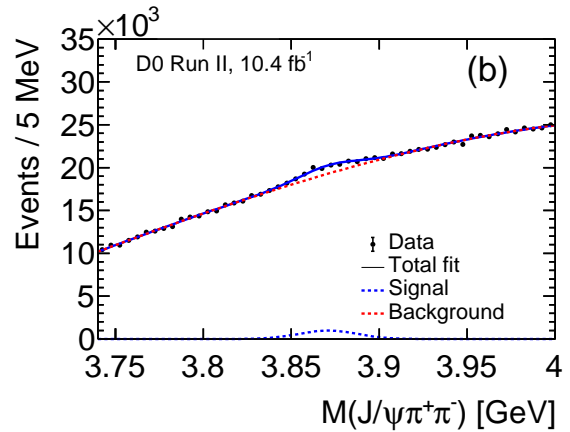}
\caption{\label{fig:mxfit5}
 The invariant mass  $M(J/\psi \pi^+  \pi^-)$
for    (a) the $\psi(2S)$ and  (b)
the $X(3872)$  selection criteria for the 5-track selection.
 }
\end{figure}

For the 5-track sample, we require the presence of an additional charged particle
with $p_T>0.5$~GeV,  consistent with coming from the same vertex.
We assume it to be a pion and  set a mass  limit   $M(J/\psi \pi^+ \pi^+ \pi^-)<4.8$~GeV.
Charge-conjugate processes are implied throughout this article.
To further reduce  background,
we allow up to two sets of three hadronic tracks  per event,
with an additional requirement that $M(J/\psi \pi^+ \pi^-)$  be less than 4~GeV.
 With up to two accepted  $J/\psi \pi^+ \pi^-$ combinations
per set, there are up to four accepted combinations per event. 
Because tracks are  ordered  by descending $p_T$,
this procedure selects  the highest-$p_T$ tracks of each charge.
Fits to the $M(J/\psi \pi^+  \pi^-)$ distribution for the 5-track selection
are shown in Fig.~\ref{fig:mxfit5}.
The fits yield  $75\,406\pm 1435$ and  $8\,192\pm 671$ signal events of  $\psi(2S)$ and $X(3872)$.
The 5-track sample is used in the studies  presented in Section~\ref{sec:5tracks}.

\section{\label{sec:4tracks}Pseudo-proper time distributions of  {\boldmath $\psi(2S)$} and  {\boldmath $X(3872)$} }

In this section we study the pseudo-proper time distributions for the charmonium states $\psi(2S)$ and $X(3872)$ using the 4-track sample.
These states can originate from the primary $p \bar p$ interaction vertex (prompt production),
or they can originate from a displaced secondary vertex corresponding to a beauty hadron decay
(nonprompt production). The pseudo-proper time $t_{pp}$ is calculated using the formula
$t_{pp} =  \vec L_{\rm xy}\hspace{-0.07cm}\cdot\hspace{-0.05cm}\vec{p}_T \, m  / (p_T^2 \,c)$, where
$\vec{p}_T$ and $m$ are the transverse momentum and mass
of the charmonium state $\psi(2S)$ or $X(3872)$ expressed in natural units and $c$ is the speed of light.
We note that the true lifetimes of $b$ hadrons decaying to $\psi(2S)$ or $X(3872)$ mesons
are slightly different from the pseudo-proper time values obtained from the formula,
because the boost factor of the charmonium is not exactly equal to the
boost factor of the parent.
Therefore, the nonprompt pseudo-proper charmonium time distributions will have effective exponential lifetime values,
which are close to but not equal to the lifetime for an admixture of $B^0$, $B^-$, $B_s^0$, $B_c^-$ mesons, and $b$ baryons.

To obtain the $t_{pp}$ distributions, the numbers of events are extracted from fits
for the $\psi(2S)$ and $X(3872)$ signals in mass distributions.
This method removes combinatorial backgrounds and yields background-subtracted numbers of $\psi(2S)$ or $X(3872)$
signal events produced in each time interval.
The bin width of the pseudo-proper time distributions is chosen to increase exponentially to reflect the exponential shape of the lifetime distributions.

The fit function used to describe the $\psi(2S)$ mass distribution includes two terms:
a single Gaussian used to model the signal and a third-order Chebyshev polynomial used to describe
background. In the specific $p_T$ and $\eta$ intervals, the statistics in some $t_{pp}$ bins may be insufficient for the fit to converge. 
In the case of a low number of background events, a second-order or a first-order
Chebyshev polynomial is used. If the number of signal events is small, the signal Gaussian mass and width
are fixed to the central values obtained in the fit to the distribution including all accepted events.
Possible variations in the parameters appearing in this approach
are estimated and are included in the systematic uncertainty.

\begin{figure}[htbp]
\centering
\epsfig{file=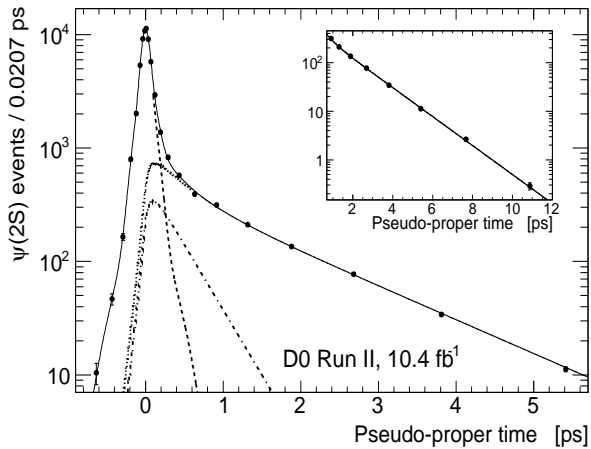,width=8.8cm,height=6.6cm}
\caption{The number of events/0.0207 ps obtained using fits to the mass distributions for the $\psi(2S)$ sample 
in pseudo-proper time bins is shown.
The tail of this distribution for the large-time region is given in the inset. The solid
curve shows the result of the fit by the function described in the text.
Also shown are contributions from the prompt component (dashed curve), the nonprompt component (dotted curve)
and the short-lived component (dash-dotted curve) of the nonprompt production.}
\label{fig:psilt}
\end{figure}

The $t_{pp}$ distribution for the $\psi(2S)$ sample is shown in Fig.~\ref{fig:psilt}.
The numbers of events/0.0207 ps shown in Fig.~\ref{fig:psilt} 
are obtained from fits to the mass distribution and corrected to the bin center to account for the steeply falling distribution.

The obtained $t_{pp}$ distributions include prompt and nonprompt contributions.
The prompt production is assumed to have a strictly zero lifetime,
whereas the nonprompt component
is assumed to be distributed exponentially starting from zero. These ideal signal distributions
are smeared by the detector vertex resolution.
The shape of the smearing function is expected to be the same for prompt and nonprompt
production.
Negative time values are possible due to the detector resolution of primary and secondary vertices.
The pseudo-proper time distribution parametrization method is similar to that used in the ATLAS analysis~\cite{8tev3872}.
For the $\psi(2S)$ sample, the $t_{pp}$ distributions are fitted using
the $\chi^2$ method with a model that includes prompt and nonprompt components:
\begin{equation}
F(t) = N \  [\ (1 - f_{NP}) \ F_P (t) + f_{NP} \ F_{NP} (t)\ ].
\end{equation}

\noindent
Here $N$ is a free normalization factor, $f_{NP}$ is a free parameter corresponding
to the nonprompt contribution fraction,
and $F_P (t)$ and $F_{NP} (t)$ are the shapes of the prompt and nonprompt components. The shape of
the prompt
component is modeled by a sum of three Gaussian functions with zero means and free normalizations and widths:
\begin{equation}
F_P (t) = g_1 G_1 + g_2 G_2 + g_3 G_3,
\end{equation}
\noindent
where $g_1$, $g_2$ and $g_3$ are normalization parameters and $G_1$, $G_2$ and $G_3$ are Gaussian functions.
The $\psi(2S)$ time distribution fit yields the three Gaussian widths $\sigma_1=0.0476\pm 0.0016$~ps,
 $\sigma_2=0.1059\pm0.0047$~ps, and $\sigma_3=0.264\pm0.
021$~ps, and
the relative normalization factors $g_1=0.491\pm0.035$, $g_2=0.447\pm0.039$, and $g_3=0.062\pm0.013$.

The shape of the nonprompt function $F_{NP} (t)$ includes two terms, a short-lived ($SL$) component and a
long-lived ($LL$) component:
\begin{equation}
F_{NP} (t)  = (1 - f_{SL}) \ F_{LL} (t) + f_{SL} \ F_{SL} (t).
\end{equation}

\noindent
The $f_{SL}$ is a free parameter in the fit. The long-lived and short-lived shape
functions $F_{LL} (t)$ and $F_{SL} (t)$ are described by single exponential functions
with slopes $\tau_{LL}$ and $\tau_{SL}$,  convolved with the resolution shape function that is the
same as for the prompt component:
\begin{equation}
F_{LL} (t) = 1 / \tau_{LL}\ \it{exp} (-\tau_{LL} \ t) \otimes F_P (t),
\end{equation}
\begin{equation}
F_{SL} (t) = 1 / \tau_{SL}\ \it{exp} (-\tau_{SL} \ t ) \otimes F_P (t).
\end{equation}

The long-lived component corresponds to charmonium
production from $B^0$, $B^+$, $B_s^0$, and other $b$ hadron decays, whereas the short-lived component
is due to the $B_c^+$ decays.
The production rate of the $B_c^+$ mesons in the $p\bar{p}$ collisions at 1.96~TeV is
not well known.
Theoretically, the ratio of $B_c^+$ meson production over all $b$ hadrons
is expected to be about ($0.1$--$\,0.2)\,\%$~\cite{pdg}.
However, the production ratio of $B_c^+$ to $B^+$ mesons has been measured
by CDF~\cite{cdfcs}, and an unexpectedly large value for this ratio between 0.9\,$\%$ and 1.9\,$\%$ was obtained;
this ratio was calculated using theoretical
predictions for the branching fraction ${\cal B}(B_c^+ \to J/\psi \mu^+ \nu)$ to be
in the range ($1.15$--$\,2.37)\,\%$~\cite{cdfcs}.
Assuming that the $\psi(2S)$ production rate in $B_c^+$ decays is enhanced by a factor of $\sim\,$20
compared with $B^+$, $B^0$ and $B_s^0$ decays,
we expect a value of $f_{SL}$ in the range of about 0.08--\,0.15.
This factor can be estimated by taking into account that the $B_c^+$ meson decays to charmonium states via the dominant ``tree'' diagram, 
whereas other $B$ hadrons produce charmonium via the ``color-suppressed'' diagram.
On the other hand, the short-lived component $f_{SL}$ was measured by ATLAS~\cite{8tev3872} in $pp$ collisions at the center-of-mass
energy 8~TeV, and a value of a few percent was obtained
for $\psi(2S)$, and one of $0.25 \pm 0.13 \pm 0.05$ for $X(3872)$.
Because of the range of possible values, we include the short-lived term with a free normalization
in the lifetime fit for the $\psi(2S)$ sample.

The $t_{pp}$ distribution of the $\psi(2S)$ sample shown in Fig.~\ref{fig:psilt} is well described
by the function discussed above, where
the exponential dependence is clearly seen in the large-time region.
The fit quality is reasonably good, $\chi^2 / {\rm NDF} = 24.5 / 14$,
corresponding  to a $p$-value of 4$\,\%$.
This fit quality is adequate in view of the large range of numbers of events per bin and the simplicity
of the pseudo-proper time fitting function.
The fitted value of the short-lived component is $f_{SL} = 0.218 \pm 0.025$.
If the short-lived component is neglected, a significantly larger value of
$\chi^2 = 112$ is obtained.
The parameters obtained from the fit shown in Fig.~\ref{fig:psilt} are listed in Table~\ref{tab:bfrpsi2s}.

\renewcommand{\arraystretch}{1.3}
\begin{table}[htb]
\vspace{-0.3cm}
\caption{The parameters obtained from the $\psi(2S)$ sample fit shown in Fig.\ref{fig:psilt}.}
\begin{center}
\vspace{-0.1cm}
\label{tab:bfrpsi2s}
\begin{tabular}
{@{\hspace{0.3cm}}l@{\hspace{0.3cm}} @{\hspace{0.3cm}}c@{\hspace{0.3cm}}}
\hline \hline
 Parameter & Fitted values, $\psi(2S)$  \\
\hline
$f_{NP}$ & 0.328 $\pm$ 0.006 \phantom{ps} \\
$f_{SL}$ & 0.218 $\pm$ 0.025 \phantom{ps} \\
$\tau_{LL}$ & 1.456 $\pm$ 0.026 ps \\
$\tau_{SL}$ & 0.38 $\pm$ 0.06 ps \\
\hline \hline
\end{tabular}
\end{center}
\vspace{-0.3cm}
\end{table}

\begin{figure}[htb]
\centering
\epsfig{file=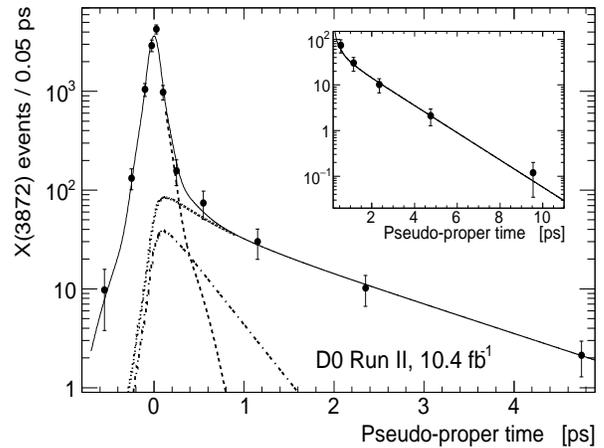,width=8.8cm,height=6.6cm}
\caption{The number of events/0.05 ps obtained using fits to the mass distributions for the $X(3872)$ sample 
in pseudo-proper time bins is shown.
The tail of this distribution for the large-time region is given in the inset. The curve shows
the result of the fit to the function described in the text.
Also shown are contributions from the prompt component (dashed curve), the nonprompt component (dotted curve)
and the short-lived component (dash-dotted curve) of the nonprompt production.}
\label{fig:xtauall}
\end{figure}

A similar method is used to obtain the pseudo-proper time distribution for the $X(3872)$ sample.
The numbers of events/0.05 ps are shown in Fig.~\ref{fig:xtauall}.
Because the number of $X(3872)$ events is an order of magnitude smaller and the combinatorial
background under the signal is slightly larger than for the $\psi(2S)$ sample, 
the number of $t_{pp}$ bins for the mass fits is reduced from 24 to 12.
The following assumptions are applied in the fit procedure: the vertex reconstruction resolution
is the same for the $X(3872)$ and $\psi(2S)$ states, and the short-lived and long-lived component lifetimes and
relative rates are fixed for the $X(3872)$ to the values obtained from the $\psi(2S)$ fit.
These assumptions are based on similarity in production kinematics and an only 5$\,\%$ difference
in the masses of these states. The relative short-lived and long-lived rates are expected to be similar,
if the ratio of inclusive branching fractions from the $B_c^+$ and other $B$ hadrons is similar for the $X(3872)$ and $\psi(2S)$ states.
The uncertainties of these assumptions are estimated and included in systematics. 
These systematic uncertainties are significantly smaller than the statistical uncertainties,
because the $f_{NP}$ values for $X(3872)$ are small and the statistical uncertainties are large.
Therefore, in the $X(3872)$ $t_{pp}$ fit procedure, all parameters are fixed to the values obtained in
the $\psi(2S)$ pseudo-proper time fit, except the $f_{NP}$ parameter.
The prompt signal Gaussian widths are scaled by the mass ratio $M(X(3872))$/$M(\psi(2S))$ to correct
for the difference in the boost factors of the $X(3872)$ sample relative to the $\psi(2S)$ sample,
which results in a different time resolution for the same spatial resolution.
We obtain $f_{NP}$ = 0.139 $\pm$ 0.025 from the fit with $\chi^2$\,/\,NDF = 8.1\,/\,10.

The systematic uncertainties on $f_{NP}$ estimated for the full $p_T$ region are listed in Table~\ref{tab:bfrx}.
They include the uncertainty due to (1) the muon reconstruction and
identification efficiencies, (2) variation of the pion reconstruction efficiency in the low- and high-$t_{pp}$ regions, 
(3) different $p_T$ distribution shapes for the prompt and nonprompt events,
(4) variation of the mass fit model parameters,
(5) variation of the time resolution function, (6) variation of the short-lived function shape,
(7) variation of the long-lived function shape, and (8) production ratio of the short-lived and long-lived components.

For the full $p_T$ range studied, we obtain $f_{NP}$ = \mbox{0.328 $\pm$ 0.006 $^{+0.010}_{-0.013}$} for the $\psi(2S)$ meson sample
and $f_{NP}$ = 0.139 $\pm$ 0.025 $\pm 0.009$ for the $X(3872)$ meson sample, where the first
uncertainty is statistical and the second is systematic.

\renewcommand{\arraystretch}{1.4}

\begin{table}[htb]
\vspace{-0.3cm}
\caption{The systematic uncertainties in $f_{NP}$ (in percent) of the $\psi(2S)$ and $X(3872)$ states.}

\begin{center}
\vspace{-0.1cm}
\label{tab:bfrx}
\begin{tabular}
{@{\hspace{0.1cm}}l@{\hspace{0.2cm}} @{\hspace{0.2cm}}c@{\hspace{0.2cm}} @{\hspace{0.2cm}}c@{\hspace{0.2cm}}}
\hline \hline
 Parameter & $\psi(2S)$ & $X(3872)$ \\
\hline
Muon reconstruction/ID efficiency & $\pm$ 0.1 & $\pm$ 0.1 \\
Pion reconstruction efficiency & $_{-0.3}^{+0.7}$ & $_{-0.2}^{+0.4}$ \\
$p_T$ distributions & $\pm$ 0.3 & $\pm$ 0.2 \\
Mass fit model & $_{-1.0}^{+0.5}$ & $_{-0.7}^{+0.5}$ \\
Resolution function & $\pm$ 0.1 & $\pm$ 0.1 \\
Short-lived (SL)component shape & $\pm$ 0.3 & $_{-0.2}^{+0.3}$ \\
Long-lived (LL) component shape & $\pm$ 0.2 & $_{-0.2}^{+0.3}$ \\
Ratio of LL and SL components & $_{-0.5}^{+0.1}$ & $_{-0.4}^{+0.5}$ \\
\hline
Sum & $_{-1.3}^{+1.0}$ & $\pm$0.9 \\
\hline \hline
\end{tabular}
\end{center}
\vspace{-0.1cm}
\end{table}

\begin{figure}[htb]
\centering
\epsfig{file=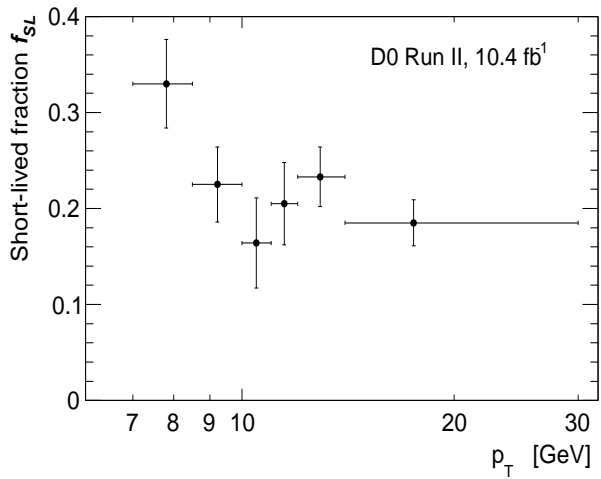,width=8.8cm,height=6.6cm}
\caption{The short-lived component fraction $f_{SL}$ as a function of $p_T$ for the $\psi(2S)$ states.
Only statistical uncertainties are shown.}
\label{fig:fsl}
\end{figure}

The large sample sizes allow us to study the $t_{pp}$ distributions in several $p_T$ intervals. We choose six
$p_T$ intervals for the $\psi(2S)$ and three for the $X(3872)$. In addition, the fit procedure is performed by dividing
the full data samples into two $\psi(2S)$ and $X(3872)$ pseudorapidity intervals:
$\left| \eta \right|$ $<$ 1 and 1 $<$ $\left|\eta \right|$ $<$ 2. The method used
to obtain parameters is the same as for the full data sample. For a given $p_T$ or $\eta$ interval,
we first fit the $\psi(2S)$ $t_{pp}$ distribution and obtain the free parameters. Then, these parameters
are fixed in the fit of the $X(3872)$ $t_{pp}$ distribution. For both mesons, the fraction $f_{NP}$ of the
nonprompt component is allowed to vary in each $p_T$ or $\eta$ interval.
Figure~\ref{fig:fsl} shows the $p_T$ dependence of $f_{SL}$ for the $\psi(2S)$;
the values of this parameter are larger than the values of
a few percent obtained by the ATLAS Collaboration~\cite{8tev3872}.

\begin{figure}[htb]
\centering
\epsfig{file=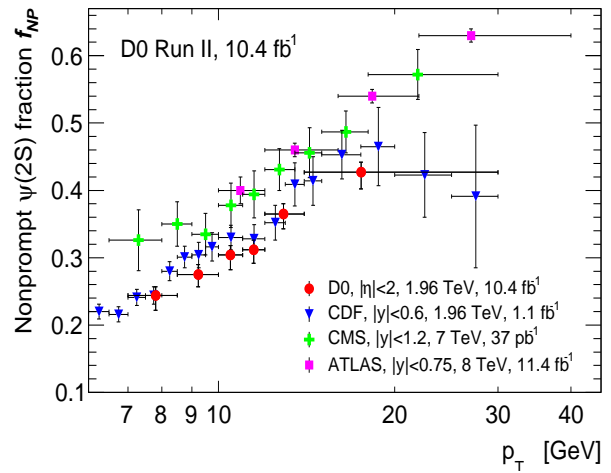,width=8.8cm,height=6.6cm}
\caption{The nonprompt component $f_{NP}$ for the $\psi(2S)$ states as a function of $p_T$.
Red circles
correspond to this analysis, magenta boxes to the ATLAS~\cite{8tev3872} measurement, green crosses to
 the CMS~\cite{cms2} measurement, and blue triangles to CDF~\cite{cdffb}.
The uncertainties shown are total uncertainties, except for the CDF points, for which only the statistical
 uncertainties
are displayed.
The D0 and ATLAS analyses are performed using $\psi(2S) \to J/\psi \pi^+ \pi^-$ decay channel,
 whereas the CMS and CDF data are obtained
through the $\psi(2S)\to \mu^+\mu^-$ decay.}
\label{fig:rnpj}
\end{figure}

\begin{figure}[htb]
\centering
\epsfig{file=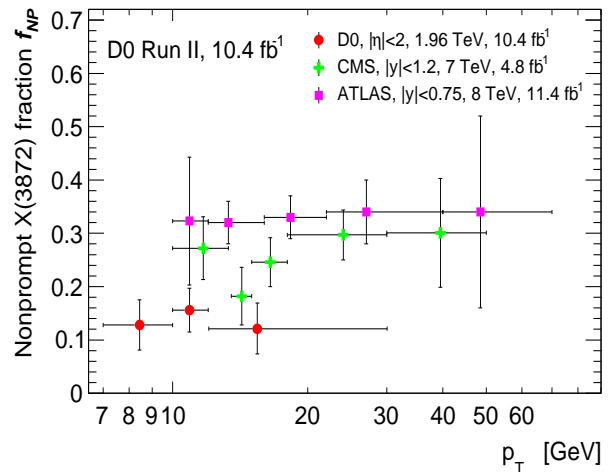,width=8.8cm,height=6.6cm}
\caption{The nonprompt component $f_{NP}$ for the $X(3872)$ states as a function of $p_T$. Red circles
correspond to this analysis, magenta boxes to the ATLAS~\cite{8tev3872} measurement and green crosses
 to the
CMS~\cite{cms} measurement. The uncertainties shown are total uncertainties.}
\label{fig:rnpx}
\end{figure}

For all measured $f_{NP}$ values the systematic uncertainties are calculated applying the same procedure and
the same variation intervals as for the whole data sample.
The values of nonprompt fractions for the $\psi(2S)$ and $X(3872)$ states in different $p_T$ or $\eta$ 
intervals
with the statistical and systematic uncertainties are given in Table~\ref{tab:bfrvspt}.
Figure~\ref{fig:rnpj} shows $f_{NP}$ as a function of $p_T$ for the $\psi(2S)$, compared with
the ATLAS~\cite{8tev3872} measurement at 8~TeV, the CMS~\cite{cms2} measurement at 7~TeV,
and the CDF~\cite{cdffb} measurement at 1.96~TeV.
Figure~\ref{fig:rnpx} shows similar distributions for the $X(3872)$ obtained in this analysis, together
 with the
ATLAS~\cite{8tev3872} and CMS~\cite{cms} measurements.
The D0 measurements of $f_{NP}$
are systematically below the ATLAS~\cite{8tev3872} and CMS~\cite{cms} points obtained at higher center-of-mass energies,
although the LHC measurements are restricted to more central pseudorapidity regions.
The small differences between the CDF and D0 $\psi(2S)$ measurements can be ascribed to differences in pseudorapidity
acceptance. 
However, the general tendencies are very similar: the $f_{NP}$ values increase with $p_T$ in the case of
 $\psi(2S)$ state
production, whereas the $f_{NP}$ values for $X(3872)$ are independent of $p_T$ within large uncertainties.

\renewcommand{\arraystretch}{1.5}
\begin{table}[htb]
\vspace{-0.3cm}
\caption{The values of nonprompt fractions $f_{NP}$ for the $\psi(2S)$ and $X(3872)$ states
in $p_T$ and $\eta$ intervals
with the statistical and systematic uncertainties are given.}
\begin{center}
\vspace{-0.1cm}
\label{tab:bfrvspt}
\begin{tabular}
{@{\hspace{0.1cm}}l@{\hspace{0.01cm}}@{\hspace{0.01cm}}c@{\hspace{0.1cm}} @{\hspace{0.3cm}}l@{\hspace
{0.01cm}} @{\hspace{0.01cm}}c@{\hspace{0.01cm}}}
\hline \hline
 & $\psi(2S)$ &  & $X(3872)$ \\
\hline
All & 0.328\,$\pm$\,0.006\,$^{+0.010}_{-0.013}$ &  & 0.139\,$\pm$\,0.025\,$\pm$ 0.009 \\
\hline
$p_T$, GeV &  & $p_T$, GeV & \\
\hline
7\,-\,8.5 & 0.244\,$\pm$\,0.008\,$^{+0.010}_{-0.021}$ & 7\,-\,10 & 0.128\,$\pm$\,0.046\,$^{+0.009}_{-0.008}$ \\
8.5\,-\,10 & 0.275\,$\pm$\,0.007\,$^{+0.013}_{-0.016}$ & & \\
10\,-\,11 & 0.304\,$\pm$\,0.009\,$^{+0.011}_{-0.020}$ & 10\,-\,12 & 0.156\,$\pm$\,0.038\,$^{+0.016}_{-0.014}$ \\
11\,-\,12 & 0.312\,$\pm$\,0.010\,$^{+0.010}_{-0.017}$ & & \\
12\,-\,14 & 0.365\,$\pm$\,0.008\,$^{+0.013}_{-0.021}$ & 12\,-\,30 & 0.121\,$\pm$ 0.047\,$^{+0.010}_{-0.006}$ \\
14\,-\,30 & 0.427\,$\pm$\,0.007\,$^{+0.013}_{-0.024}$ & & \\
\hline
\end{tabular}
\begin{tabular}
{@{\hspace{0.1cm}}l@{\hspace{0.3cm}}@{\hspace{0.4cm}}c@{\hspace{0.4cm}} @{\hspace{0.3cm}}c@{\hspace{0.1cm}}}
 & $\psi(2S)$ &  $X(3872)$ \\
\hline
$\left|\eta\right|<1$ & 0.344\,$\pm$\,0.007\,$^{+0.014}_{-0.020}$ &  0.164\,$\pm$\,0.035\,$^{+0.009}_{-0.016}$ \\
$1<\left|\eta\right|<2$ & 0.303\,$\pm$\,0.008\,$^{+0.017}_{-0.020}$ &  0.116\,$\pm$\,0.032\,$^{+0.009}_{-0.010}$ \\
\hline \hline
\end{tabular}
\end{center}
\vspace{-0.1cm}
\end{table}

We summarize the measurements of this section as follows:

\begin{enumerate}
\item  The nonprompt fractions for $\psi(2S)$ increase as a function of $p_T$, whereas those for $X(3872)$ are consistent
with being independent of $p_T$. These trends are similar to those seen at the  LHC.
The Tevatron values tend to be somewhat smaller than those measured by ATLAS and CMS.

\item The ratio of prompt to nonprompt 
$\psi(2S)$ production, $R_{p/np} = (1 - f_{NP}) / f_{NP}$,
increases only slightly going from the LHC to the Tevatron. As can be seen in Fig.\,\,6, the
$f_{NP}$ values in the \mbox{9\,--10 GeV} range are
0.35$\,\pm\,$0.03 for LHC data and 0.30$\,\pm\,$0.02 for Tevatron data, 
resulting in increase in $R_{p/np}$ of (7\,--\,47)$\,\%$ (68.3$\,\%$ confidence interval).
At low $p_T$, the CMS data points have large statistical uncertainties,
but the Tevatron data can be compared to the LHCb measurement of $\psi(2S)$ $f_{NP}$ values~[18] at 7 TeV
for 2.0 $< y <$ 2.5 and \mbox{6 $< p_T <$ 14 GeV}. The LHCb $R_{p/np}$ values are about (25\,--\,30)\,$\%$ smaller
than those from the Tevatron,
after adjustment for the variation with pseudorapidity.
The LHCb data indicate a reduction of $f_{NP}$ by 0.02\,--\,0.03 for each reduction in rapidity by one unit.

\item The ATLAS value of $f_{NP}$ = 0.328 $\pm$ 0.026 for the $X(3872)$ differs from the D0 value of
$f_{NP}$ = 0.139 $\pm$ 0.027 by 5.0$\,\sigma$, taking into account both statistical and systematic uncertainties
and assuming a uniform $p_T$ distribution. This gives an increase in the $R_{p/np}$ ratio by a factor of $\sim\,$3
(the range 2.4\,--\,4.0 for the 68.3$\,\%$ confidence interval) going from the LHC to the Tevatron.
It has to be noted that this difference may be partially compensated by the larger rapidity interval covered by D0.
This increase of the $R_{p/np}$ value indicates that the prompt production of the exotic state $X(3872)$ relative
to the $b$ hadron production is strongly suppressed at the LHC in comparison with the Tevatron conditions.
This suppression is possibly due to more particles produced in the primary collision at LHC that increase the probability
to disassociate the nearly unbound and possibly spatially extended $X(3872)$~\cite{diss,lhcbiso}.

\end{enumerate}


\section{\label{sec:5tracks}Hadronic activity around the {\boldmath $\psi(2S)$ and $X(3872)$ states}}

In this section, we study the association of the  $\psi(2S)$ or  $X(3872)$ states  with
another particle assumed to be a pion using the 5-track sample.
We study  the dependence of the  production of these two states
on the surrounding hadronic activity.
We also test the soft-pion signature of the $X(3872)$ as a weakly bound
charm-meson pair by studying the production of $X(3872)$ at small
 kinetic  energy of the $X(3872)$ and the $\pi$ in the  $X(3872)\pi$ center-of-mass frame.

The data are separated into
a ``prompt'' sample, defined by the conditions $L_{\rm xy}<0.025$~cm and  $L_{\rm xy}/\sigma(L_{xy})<3$
and a ``nonprompt'' sample defined by $L_{xy}>0.025$~cm and  $L_{xy}/\sigma(L_{xy})>3$,
where  $L_{\rm xy}$ is the decay length of the  $J/\psi \pi^+ \pi^-$ system in the transverse plane.

In these studies, the uncertainties in the results
are dominated by the statistical uncertainties in the fitted $X(3872)$ yields.
 In the limited mass range around
the  $\psi(2S)$ or $X(3872)$, the background is smooth and monotonic,
and  is well
described by low-order Chebyshev polynomials. Depending on the 
size of a given subsample, the polynomial order is set to 2 or 3.
 In all cases, the difference between
the yields for the two background choices is less than 30\% of the statistical uncertainty.
The small systematic uncertainties are ignored.

\subsection{ \boldmath{$\psi(2S)$} and \boldmath{$X(3872)$}  isolation }

The LHCb Collaboration has studied~\cite{lhcbiso}  the dependence of production cross sections of
the $X(3872)$ and $\psi(2S)$ on the hadronic activity in an event,
which is approximated using a measure of the charged particle multiplicity.
The authors found the ratio of the cross sections
for promptly produced particles,  $\sigma(X(3872))/\sigma(\psi(2S))$, to decrease with increasing multiplicity
and observed that this behavior is consistent with the interpretation of the  $X(3872)$
as a weakly bound state, such as a $D^0 \bar D^{*0}$ hadronic molecule. In this scenario, interactions
with comoving hadrons produced in the collision disassociate the large, weakly bound $X(3872)$ state
more than the relatively compact conventional charmonium state $\psi(2S)$.

\begin{figure}[htb]
\includegraphics[scale=0.4]{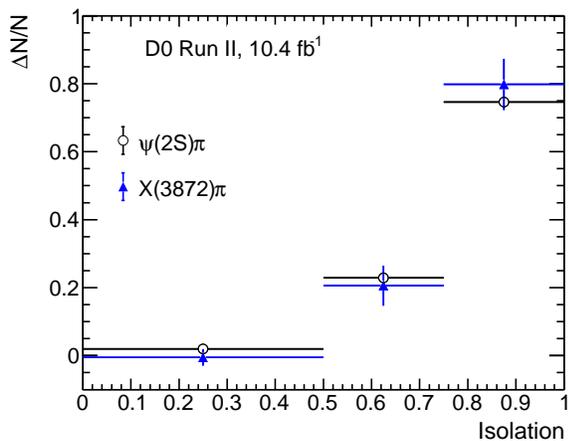}
\caption{\label{fig:xvsiso} Normalized yields of the    $\psi(2S)\pi$ (black open circles) and
the $X(3872)\pi$ (blue triangles) as functions of isolation for the prompt sample.
}
\end{figure}

\begin{figure}[htb]
\includegraphics[scale=0.4]{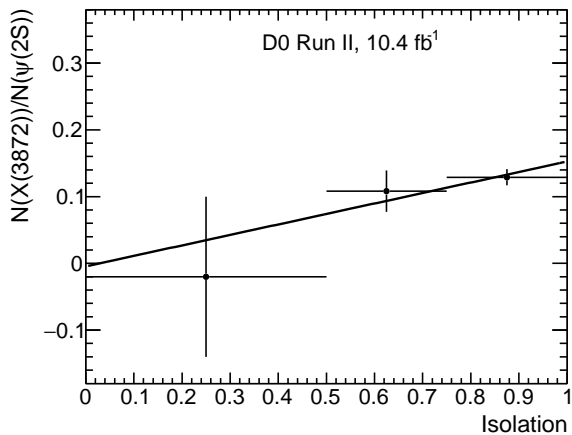}
\caption{\label{fig:rvsiso} The ratio of the unnormalized $X(3872)$ and $\psi(2S)$ yields as a function
of isolation for the prompt sample.
}
\end{figure}

In this study of the production  of charmonium-like states, we
introduce {\it isolation} as an observable quantifying the hadronic activity in a restricted  cone
in the $\phi-\eta$ space around the candidate, $\Delta R = \sqrt{{\Delta \phi}^2 + {\Delta \eta}^2}$.
We define the isolation  as
a ratio of  the  candidate's  momentum to the scalar sum of the momenta of
all charged particles pointing to the primary vertex
produced  in a cone  of  $\Delta R=1$ around the  candidate and the candidate itself.
Distributions of isolation for  prompt  $\psi(2S)\pi$ and  $X(3872)\pi$ normalized to unity
are shown in Fig.~\ref{fig:xvsiso}, and the ratio of the unnormalized 
distributions is shown in Fig.~\ref{fig:rvsiso}.
The shapes of the two isolation distributions are similar.
 The difference between the $\chi^2$ values obtained for 
 fits to the ratio as a function of isolation
 assuming  a free slope and  zero slope
corresponds to 1.2$\,\sigma$. This gives 
 modest support for the hypothesis that increased hadronic activity near $X(3872)$ depresses its production.

\subsection{Search for the soft-pion effect}

Recent theoretical work~\cite{bra3872, bra3872b}
predicts a sizable contribution to the  production of the $X(3872)$, both directly in
the hadronic beam collisions and in $b$ hadron decays,
from the formation of the $X(3872)$ in association with a comoving pion.
According to the authors, the  $X(3872)$, assumed to be a $D \bar D^*$ molecule,
is produced by the creation of $D \bar D^*$
at short distances. But it can also be produced by the creation of $D^*\bar D^*$
at short distances, followed by a rescattering of the charm-meson pair
into a  $X(3872)\pi$ pair by exchanging a $D$ meson.
The cross section from this mechanism would have  a narrow peak
in the  $X(3872)\pi$ invariant mass distribution near the  $D^*\bar D^*$ threshold
from a triangle singularity that occurs when the three particles participating in
a rescattering are all near the mass shell.

 A convenient variable to quantify this effect
is the kinetic energy $T(X\pi)$ of the $X(3872)$ and the $\pi$ in the  $X(3872)\pi$
center-of-mass frame. The authors define the peak region to be
$0 \leq  T(X\pi) \leq 2\delta_1$ where $\delta_1= M(D^{*+}) - M(D^0) - M(\pi^+) =5.9$~MeV.
The  effect is sensitive to the $D \bar D^*$
binding energy whose current estimated value is $(-0.01\pm 0.18)$~MeV. The peak height is expected to decrease
with increasing binding energy.
It also depends on the value of the momentum scale $\Lambda$, expected to be of the order of $M(\pi^+)$.
For the conservative choice of a binding energy of 0.17~MeV, the yield in the peak region is predicted to  be smaller
 than the  yield without a soft pion by a factor
  $\sim 0.14(M(\pi^+)/\Lambda)^2$.  For $\Lambda=M(\pi^+)$, this ratio
is equal to  0.14.
We search for this effect separately in the  ``prompt'' and ``nonprompt''
samples.

\vspace{-5.mm}
\subsubsection*{{\textbf {\textit 1. Prompt production}}}
\vspace{-2.mm}

\begin{figure}[htb]
\includegraphics[scale=0.4]{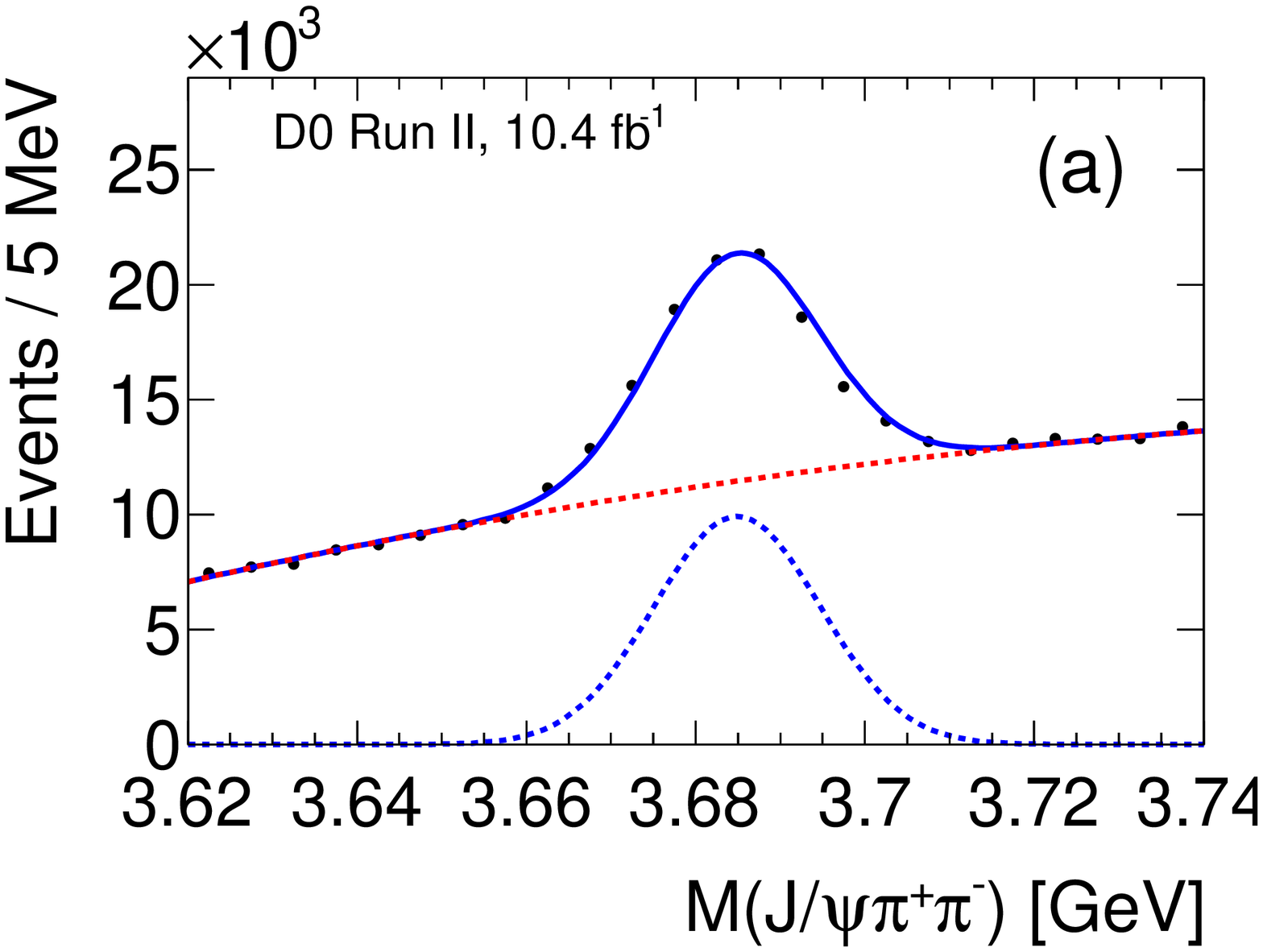}
\includegraphics[scale=0.4]{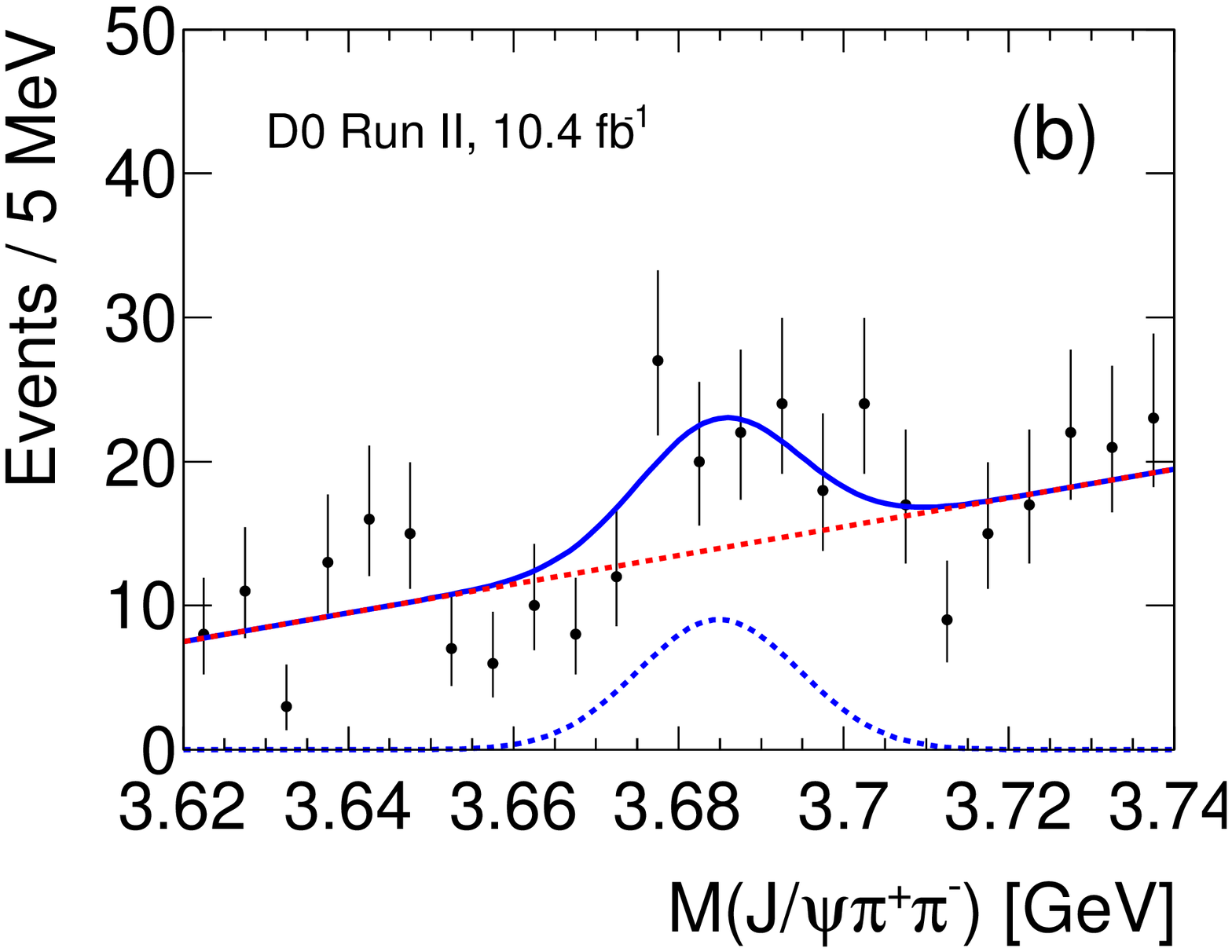}
\caption{\label{fig:mpsipie} $M(J/\psi \pi^+  \pi^-)$ distribution and fits for
the $\psi(2S)$ signal for the prompt subsample for
(a) all selected events and (b)  events passing the $T(\psi(2S)\pi)<11.8$~MeV cut.
}
\end{figure}

As a benchmark, we use  the $\psi(2S)$, for which no soft-pion effect is expected. We select
 combinations $J/\psi \pi^+ \pi^+ \pi^-$
that have  a $J/\psi \pi^+  \pi^-$  combination in the mass range
$3.62<M(J/\psi \pi^+ \pi^-)<3.74$ GeV.
The total number of entries is 310\,636, and the  $\psi(2S)$ signal
has  $48\,711\,\pm\,511$ events.
The mass distributions and fits are shown in Fig.~\ref{fig:mpsipie}.
After the  $T(\psi(2S)\pi)<11.8$~MeV cut, the number of entries is
368 and the signal yield is $44\pm14$ events.
The cut  $T(\psi(2S)\pi)<11.8$~MeV keeps a fraction $0.0009\pm0.0003$ of the signal,
in agreement with the measured reduction of the combinatorial background
by a factor of 0.0012.
As expected, there is no evidence for a soft-pion effect for $\psi(2S)$.

\begin{figure}[htb]
\includegraphics[scale=0.4]{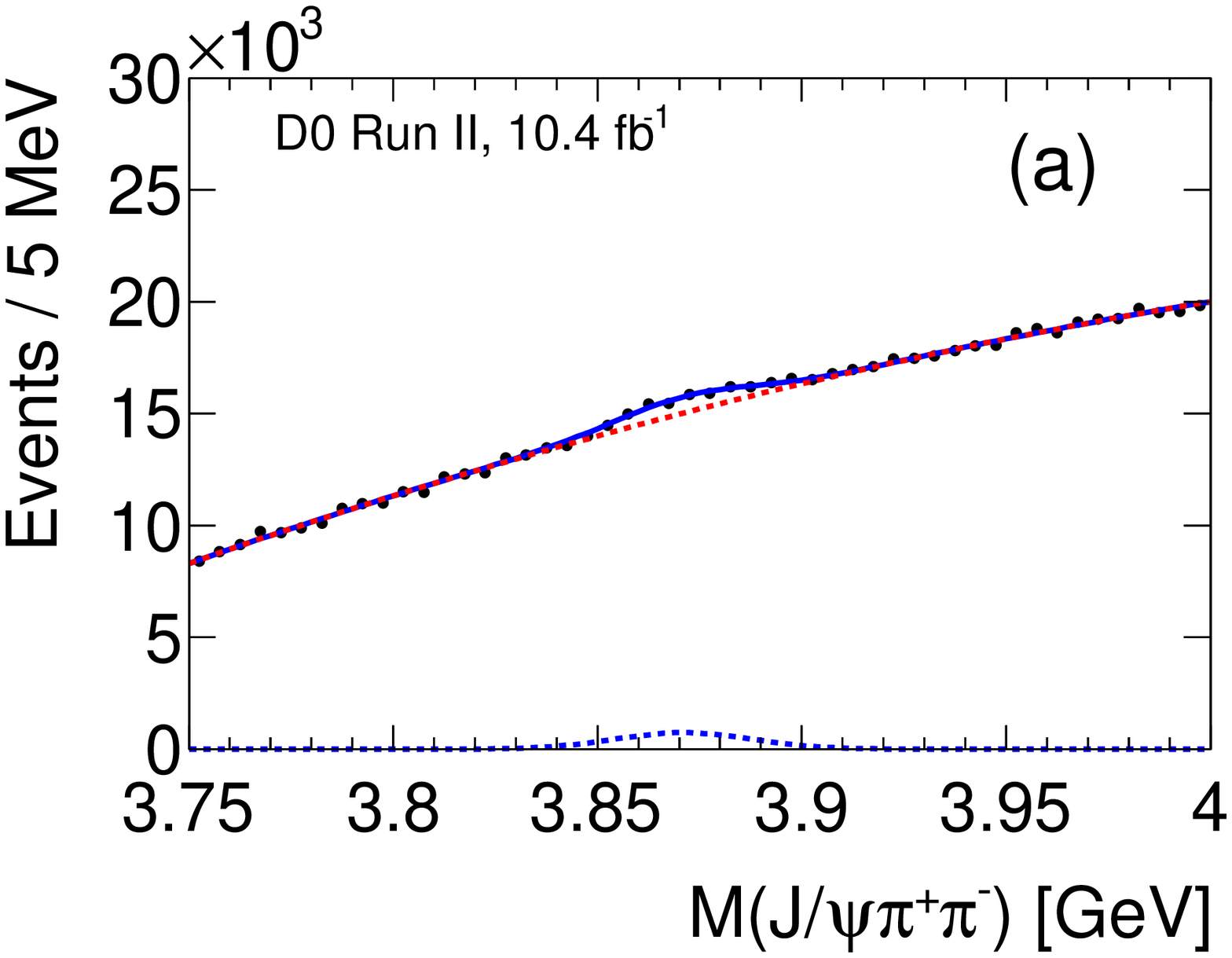}
\includegraphics[scale=0.4]{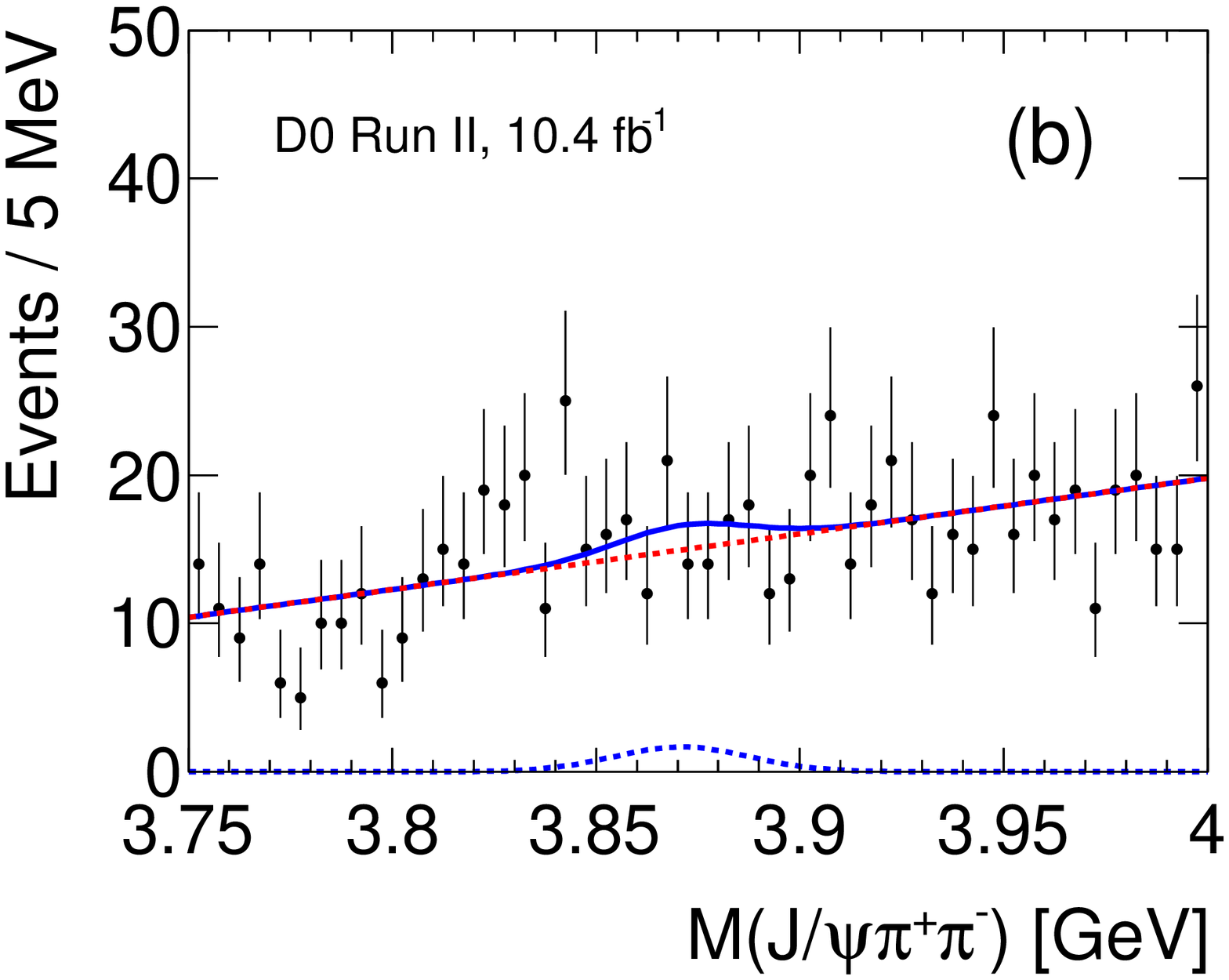}
\caption{\label{fig:mxpie} $M(J/\psi \pi^+  \pi^-)$ distribution and fits for
the $X(3872)$ signal for the prompt subsample for
(a) all selected events and (b)  events passing the $T(X\pi)<11.8$~MeV cut.
}
\end{figure}

Then, we select $J/\psi \pi^+ \pi^+ \pi^-$ combinations
that have a $J/\psi \pi^+  \pi^-$  combination in the mass range
$3.75<M(J/\psi \pi^+ \pi^-)<4.0$ GeV, that includes the $X(3872)$.
The total number of selected entries is 749\,179 and the  $X(3872)$ signal yield is
 $6\,157\pm 599$ events.
The mass distributions and fits are shown in Fig.~\ref{fig:mxpie}.
The signal consists of a $X(3872)$ meson produced together with a charged particle.
It includes possible pairs of a  $X(3872)$ meson and an associated soft pion
from the triangle singularity. The background is due to  random combinations
of a $J/\psi$ meson and three charged particles.
The cut  $T(X\pi)<11.8$ MeV should remove the bulk of random   $X(3872)$-pion
combinations while keeping the events due to the triangle singularity.
For this subsample of 730 events, the fitted signal yield is $18\pm16$ events.
Thus, the cut  $T(X\pi)<11.8$~MeV keeps a fraction $0.003\pm0.003$ of the signal,
 consistent with the background
reduction by a factor of 0.00097$\,\pm\,$0.00004. In the absence of the soft-pion process,
the expected yield at small $T(X\pi)$ is $N=6157\times 0.00097=6$ events. With the measured
yield of  $18\pm16$ events, the net excess is 12$\,\pm\,$16 events. The 90\% C.L.
upper limit is 43 events,
which is  less than 0.007 of the total number of accepted events.

To compare this result with the expected number of accepted soft-pion events,
we make a rough estimate of the kinematic acceptance for events above and below the 11.8~MeV cutoff.
The main factor is the loss of pions produced with $p_T<0.5$~GeV that strongly depends on  $T(X\pi)$,
given the $p_T$ distribution of the $X(3872)$.

\begin{figure}[htb]
\includegraphics[scale=0.4]{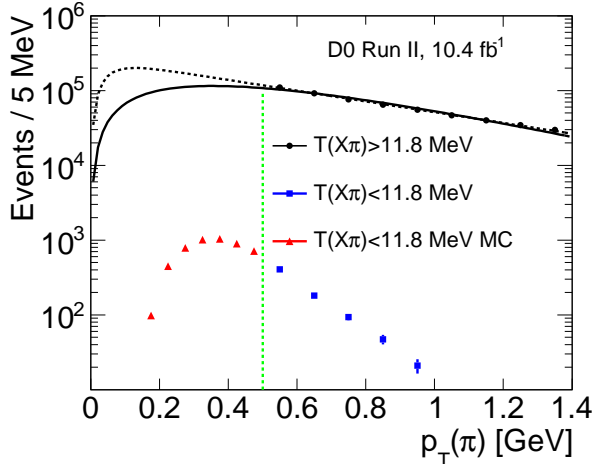}
\caption{\label{fig:ptpi}
Transverse momentum distribution of pion candidates for events above and below the $T(X\pi)=11.8$~MeV cutoff
for prompt events in the mass range  $3.75<M(J/\psi \pi^+ \pi^-)<4$~GeV.
The former is compared to two fits discussed in the text.
Extrapolation of the latter below threshold follows the method described in the text.
}
\end{figure}

The transverse momentum distributions of pions in the two subsamples are shown in Fig.~\ref{fig:ptpi}.
Above 0.5~GeV,  the distributions fall exponentially.
Below the 0.5~GeV threshold, the spectrum must rise from  the minimum kinematically allowed value  to a peak followed by
 the exponential falloff.
For events with   $T(X\pi)>$11.8~MeV,
we fit the distribution  to the function $N\cdot p_T \cdot \exp(-p_T/p_{T0})$  
and define the acceptance $A$ as
the ratio of the integral from 0.5~GeV to infinity to the integral from zero  to infinity.
The result is  0.6. 
With alternate functions, the acceptance values  vary  from 0.3 to 0.9.
Figure ~\ref{fig:ptpi} shows two fits with similar behavior above threshold but
different below threshold, the default function and the function
$N\cdot (1-\exp(-p_T/p_{T1}))\cdot \exp(-p_T/p_{T2})$.

For  events  with   $T(X\pi)<11.8$~MeV, the $p_T$ distribution of the accompanying 
pion is closely related to the  $p_T$  of the $X(3872)$.
To determine the pion acceptance, we employ a  simplified MC model,
 starting with the
differential cross section as a function of  $T(X\pi)<$11.8~MeV given in Ref.~\cite{bra3872}.
For a  $X(3872)$  with a given $p_T(X)$, the $X$ and pion are distributed isotropically
in the $X\pi$ rest frame. The transverse momentum of the pion
$p_T(\pi)$ in the laboratory frame is determined by transforming to the $X(3872)$ rest frame,
using the chosen $p_T(X)$ and a rapidity $y(X)$ chosen from a uniform
distribution $\left|y\right|<2$, and then transforming to the laboratory frame.
 The pion acceptance as a function of $p_T(X)$, $A(p_T(X))$,
is then  convolved with the fitted $X(3872)$ yield $dN/dp_T(X)$ as a function
of $p_T(X)$ to determine the overall
acceptance for pions.

\begin{figure}[htb]
\includegraphics[scale=0.4]{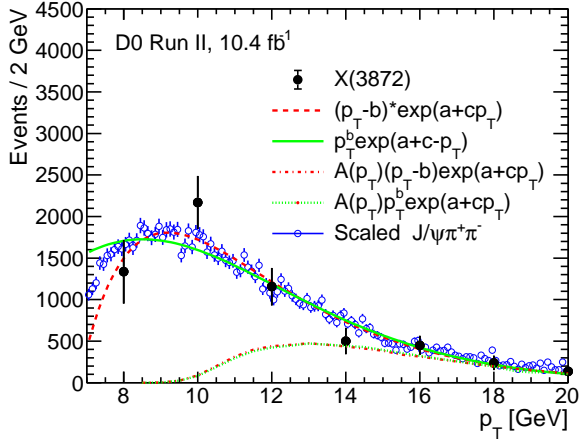}
\caption{\label{fig:Api} The transverse momentum distribution for the background-subtracted mass-fitted
 $X(3872)$ (filled circles) and two fits representing the high and low ranges of the acceptance
 for the accompanying pion.  The dashed curves represent $A(p_T) \cdot dN/dp_T(X(3872))$.  The overall
acceptance for the accompanying pion is the ratio of the areas below $A \cdot dN/p_T(X)$ curves and the
 corresponding $dN/dp_T(X)$ fits. For
comparison, the scaled $p_T$ distribution of the inclusive $J/\psi \pi \pi$ for
 $3.85<M(J/\psi \pi^+ \pi^-)<3.9$~GeV (open blue circles)
is overlaid, illustrating their similarity in shape.}
\end{figure}

Our observed $dN/dp_T$ distribution for the $X(3872)$ is found by dividing the mass distribution for
 $J/\psi \pi^+ \pi^-$ in
Fig.~\ref{fig:mxpie}(a) for the 5-track sample into seven $p_T$ bins each \mbox{2~GeV} wide, between 7 
and 21 GeV, and fitting for the yield of the
$X(3872)$ for each bin.  This produces a background-subtracted sample; however, it has relatively large
statistical uncertainties.
These seven $dN/dp_T$ yield points for the $X(3872)$ are plotted in Fig.~\ref{fig:Api}.
The higher statistics and finer-binned yield for inclusive $J/\psi \pi^+ \pi^-$ events over the mass range
3850\,--\,3900 MeV of
Fig.~\ref{fig:mxpie}(a) as a
function of $p_T$ is  used to check the shape of the $p_T$ distribution of the $X(3872)$. 
After scaling to equal areas,  $dN/dp_T(J/\psi \pi^+ \pi^-)$
 shows a good agreement within statistical uncertainties with the 
$X(3872)$ spectrum, thus indicating a comparable behavior of the $X(3872)$
 signal and background.

Fits of the background-subtracted yields using the functions $p_T^b \cdot \exp(a + c \cdot p_T)$ and
$(p_T -b) \cdot \exp(a + c \cdot p_T)$
are shown in Fig.~\ref{fig:Api}, along with the products $A(p_T) \cdot dN/dp_T$, which allow the
calculation of the 
acceptance for $p_T(\pi) >$ 0.5 GeV for events with  $p_T(X)>$~7 GeV.
We find the acceptances A = 0.278$\,\pm\,$0.031 and 0.296$\,\pm\,$0.036 for the two
 functions, respectively, where the
uncertainties are due to the statistical uncertainty in the determination of the $dN/dp_T(X)$
 distribution.
Additional functions were used to fit $dN/dp_T(X)$. The aforementioned functions yield the 
lowest and highest pion acceptances  obtained
from the different forms.  Their difference is considered as the systematic uncertainty associated with
 the choice of parametrization. We average the two results to obtain
 $A=0.29\pm 0.03$ (stat) $\pm$ 0.02 (syst).

For the prompt case, this leads to the expected number  of produced $X(3872)$ events at
 $N=18/0.29+6139/0.6 \approx 10\,000$ with an uncertainty of about $\pm\,50\%$.
With $N=N_1+N_0$, where $N_1$ is the number of events with a soft pion, and the relation $N_1=0.14\cdot N_0$,
$N \approx 10\,000\times0.14/1.14 \approx1300$ events would be produced through  the soft-pion process with an uncertainty of about 650 events,
and between 245 and 730  would be accepted. That is much larger than the observed 12$\,\pm\,$16 events.
We conclude that there is no evidence for the soft-pion effect
in the prompt sample.

\vspace{-5.mm}
\subsubsection*{{\textbf {\textit 2. Nonprompt production}}}
\vspace{-2.mm}

The kinematics of the prompt and nonprompt samples are sufficiently similar to
use the acceptance derived for the prompt case for both samples.
Calculations analogous to those for the prompt case give the following results for the nonprompt sample.
For the  $\psi(2S)$, the kinetic energy cut keeps a fraction of
$0.004\pm0.001$ of the signal, in agreement with the reduction by a
factor of 0.003 of the total number of entries.

For the $X(3872)$, the signal yields before and after the cut are
$703\pm25$ and $27\pm12$, respectively. The cut accepts a fraction
$0.04\pm0.02$ of the signal. The corresponding reduction in the
total number of events in the distribution is by a factor of $0.0029\pm0.0001$.
 For a random pairing of the $X(3872)$ with a pion,
the expected yield at small $T(X\pi)$ is $N=703\times 0.0029=2$ events,
leading to a net excess of 25$\,\pm\,$12 events.
The statistical significance of the excess,
based on the $\chi^2$ difference between the fit with a free signal yield and
the fixed value of $N=2$ expected for the ``random pairing only'' case, is $2\,\sigma$.
Correcting for soft-pion acceptance, the number of produced nonprompt 
$X(3872)$ before the kinematic cut is in the range 800 -- 2000.
Assuming the ratio of 0.14 between the cross section
for production with a soft pion to the cross section for production without
a soft pion~\cite{bra3872}, we estimate the expected number of produced soft-pion
events to be in the range 100\,--\,300.
With the acceptance of 0.29$\,\pm\,$0.04,
the expected number of accepted soft-pion
events is between 30 and 90.
The measured excess yield of $25\pm12$ events is in agreement with this expectation;
however the fact that our yield agrees within $2\,\sigma$ with the null hypothesis of
no soft-pion events prevents drawing a definite conclusion.

\begin{figure}[htb]
\includegraphics[scale=0.4]{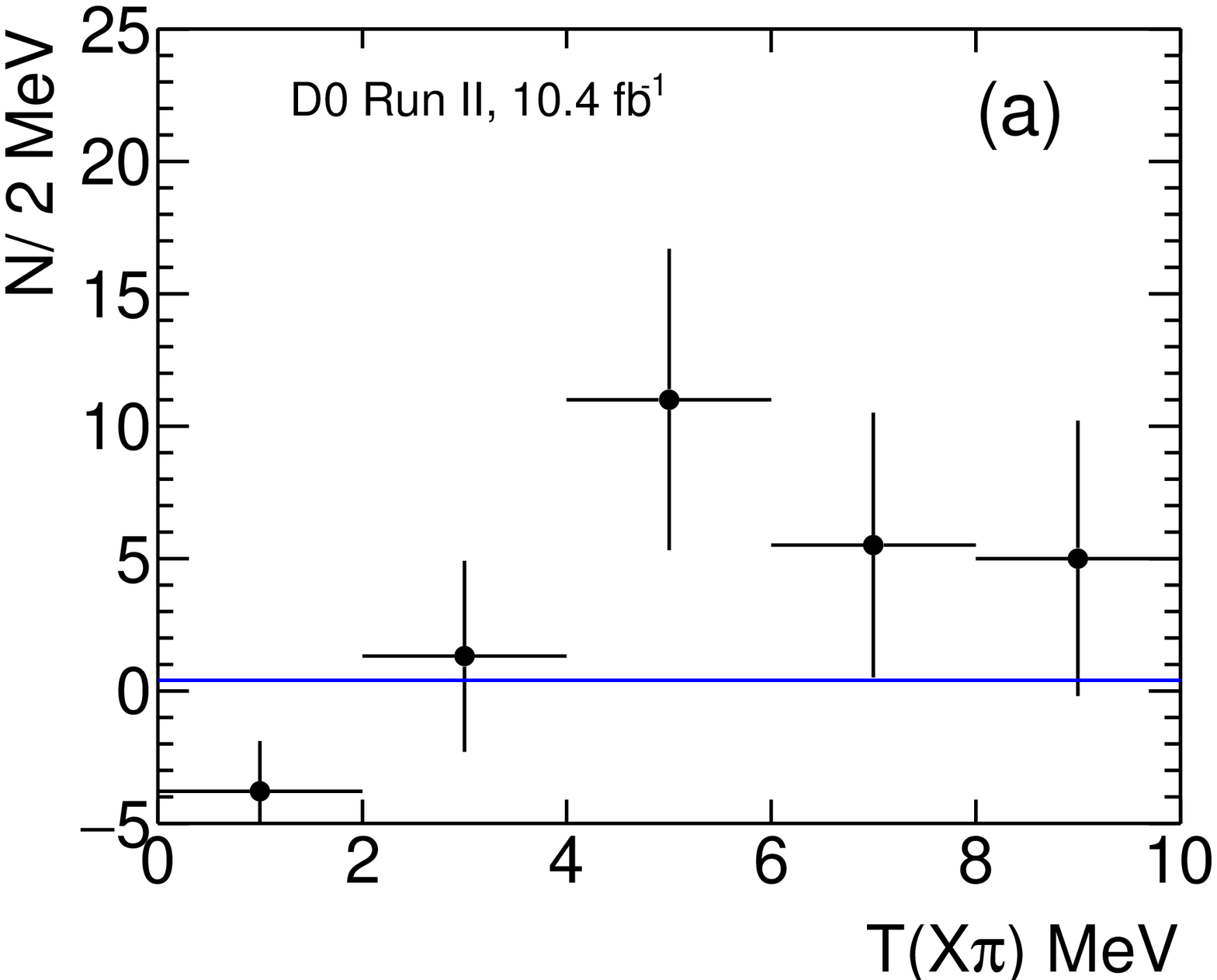}
\includegraphics[scale=0.4]{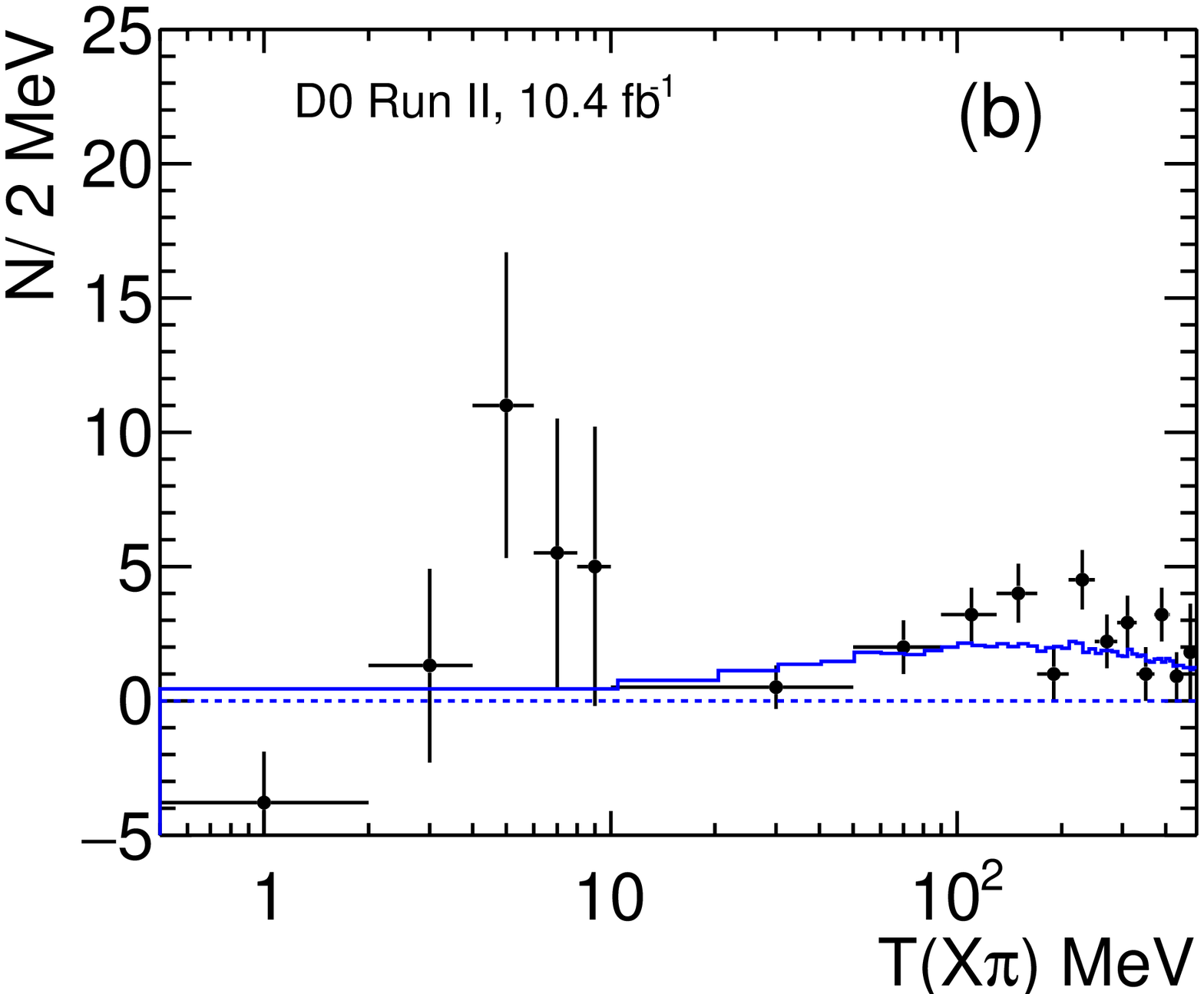}
\caption{\label{fig:xvsexpi}  The fitted $X(3872)$ signal yield as a function of  $T(X\pi)$
for nonprompt events with  (a)  the soft-pion production region and
(b) extended range. The first five points in (b) are the same as those in (a).
The blue line shows the  distribution of the $T(X\pi)$ for all nonprompt $X(3872)$ candidates
scaled down to the total $X(3872)$ yield.}
\end{figure}

For further details on the distribution of the nonprompt signal versus $T(X\pi)$, we fit the $X(3872)$ mass distributions  
into 2~MeV bins of  $T(X\pi)$
from 0 to 10~MeV and into 40~MeV bins from 10 to 490~MeV.
The resulting distribution of events/2 MeV is shown in  Fig.~\ref{fig:xvsexpi}.  Above $\sim$10~MeV, the observed spectrum
 is consistent with the pairing of a $X(3872)$ with a random
particle.  It is similar to the $T(X\pi)$ distribution of all nonprompt $X(3872)$ candidates.
At lower $T(X\pi)$, there is a small excess, with a significance of 2$\,\sigma$,  above the random pairing,
at the level consistent with the predictions of
Ref.~\cite{bra3872}.
We again conclude that there is no significant evidence for the soft-pion effect, but
  its presence at the level
expected for the binding energy of 0.17~MeV and the momentum scale $\Lambda=M(\pi)$ is not ruled out.


\section{\label{sec:sum}Summary and conclusions}

We have presented various properties  of the production of  the  $\psi(2S)$
and $X(3872)$  in Tevatron $p \bar p$ collisions.
For both states, we have measured the fraction $f_{NP}$  of the inclusive production rate
due to decays of  $b$-flavored hadrons as a function of  the  transverse momentum $p_T$.
Our nonprompt fractions for $\psi(2S)$ increase as a function of $p_T$, whereas those for $X(3872)$ are consistent
with being independent of $p_T$. These trends are similar to those seen at the  LHC.
The Tevatron values tend to be somewhat smaller than those measured by ATLAS and CMS, but this difference
can at least partially be accounted for by the larger rapidity interval covered by D0.
The ratio of prompt to nonprompt $\psi(2S)$ production, $(1 - f_{NP}) / f_{NP}$, decreases only slightly
going from the Tevatron to the LHC, 
but in comparing the 8 TeV ATLAS data to the \mbox{1.96~TeV} D0 data for the $X(3872)$ production, this ratio decreases by a factor
of approximately 3.
This indicates that the prompt production of the exotic state $X(3872)$ is suppressed at the LHC,
possibly due to the production of more particles in the primary collision that increases the probability to disassociate
the nearly unbound and possibly more spatially extended $X(3872)$ state.

We have  tested the soft-pion signature of the $X(3872)$ modeled
as a weakly bound charm-meson pair by studying the production of the $X(3872)$
as a function of the kinetic energy of the $X(3872)$ and the pion
in the  $X\pi$ center-of-mass frame.
For a subsample consistent with prompt production, the results
are incompatible with a strong enhancement in the  production of the $X(3872)$ at small  $T(X\pi)$
expected for the $X$+soft-pion  production mechanism.
For events consistent with being due to decays of $b$ hadrons,
there is no significant evidence for the soft-pion effect, but its presence at the level
expected for the binding energy of 0.17~MeV and the momentum scale $\Lambda=M(\pi)$ is not ruled out.

\section*{ACKNOWLEDGMENTS}

This document was prepared by the D0 Collaboration using the resources of the Fermi National Accelerator Laboratory (Fermilab),
a U.S. Department of Energy, Office of Science, HEP User Facility. Fermilab is managed by Fermi Research Alliance, LLC (FRA),
acting under Contract No. DE-AC02-07CH11359.

We thank Eric Braaten for useful discussions.
We thank the staffs at Fermilab and collaborating institutions,
and acknowledge support from the
Department of Energy and National Science Foundation (United States of America);
Alternative Energies and Atomic Energy Commission and
National Center for Scientific Research/National Institute of Nuclear and Particle Physics  (France);
Ministry of Education and Science of the Russian Federation, 
National Research Center ``Kurchatov Institute" of the Russian Federation, and 
Russian Foundation for Basic Research  (Russia);
National Council for the Development of Science and Technology and
Carlos Chagas Filho Foundation for the Support of Research in the State of Rio de Janeiro (Brazil);
Department of Atomic Energy and Department of Science and Technology (India);
Administrative Department of Science, Technology and Innovation (Colombia);
National Council of Science and Technology (Mexico);
National Research Foundation of Korea (Korea);
Foundation for Fundamental Research on Matter (The Netherlands);
Science and Technology Facilities Council and The Royal Society (United Kingdom);
Ministry of Education, Youth and Sports (Czech Republic);
Bundesministerium f\"{u}r Bildung und Forschung (Federal Ministry of Education and Research) and 
Deutsche Forschungsgemeinschaft (German Research Foundation) (Germany);
Science Foundation Ireland (Ireland);
Swedish Research Council (Sweden);
China Academy of Sciences and National Natural Science Foundation of China (China);
and
Ministry of Education and Science of Ukraine (Ukraine).


\begin{thebibliography}{99}
\bibitem{eta}
$\eta = --\,\ln[\tan(\theta/2)]\ $ is the pseudorapidity and $\theta$ is
the polar angle between the track momentum and the proton beam direction. $\phi$ is the azimuthal
angle of the track.
\end{thebibliography}
\end{document}